\documentclass[structabstract]{aa}  
\usepackage{graphicx}
\usepackage{txfonts}
\usepackage{natbib}
\begin{document}

\title{ Link between trees of fragmenting granules  and deep  downflows in MHD simulation}
\author{T. Roudier\inst{1}
\and
       J.M.~Malherbe\inst{2}
       \and
       R. F.  Stein \inst{3}
        \and
       Z.Frank\inst{4}       
        }
\offprints{Th. Roudier,\\
\email{thierry.roudier@irap.omp.eu}}
\institute
    {
      Institut de Recherche en Astrophysique et Plan\'etologie, Universit\'e de Toulouse, CNRS, UPS,
      CNES 14 avenue Edouard Belin, 31400 Toulouse, France
\and
Observatoire de Paris, LESIA, 5 place Janssen, 92195 Meudon, France, PSL Research University, CNRS, Sorbonne Universit\'es, UPMC Univ. Paris 06, Univ. Paris Diderot, Sorbonne Paris Cit\'e 
\and
Physics and Astronomy Department, Michigan State University, East Lansing, MI 48824, USA
\and
Lockheed Martin Solar and Astrophysics Laboratory, 3251 Hanover Street, Palo Alto, CA 94303, USA
}

\date{Received date / Accepted date }
\titlerunning{ Link between trees of fragmenting granules and deep downflows in MHD simulation}
\authorrunning{Roudier et al.}

\abstract
{Trees of fragmenting granules (TFG) and associated flows are suspected to play
a major role in the formation of the network in the quiet Sun. We investigate the counterparts, in terms of dynamics, of surface structures detectable by high resolution observations in deeper layers up to 15 Mm, which are only available from numerical simulations.}
{ The first aim is to demonstrate that TFG can be evidenced either from surface intensitites,
  vertical (Vz), or Doppler (Vdop) velocities. The second is to show that horizontal flows, which are derived
  from intensities or Vz/Vdop flows, are in good agreement, and that this is the case for observations
  and numerical simulations. The third objective is to apply this new Vz-based method to a 3D simulation
to probe relationships between horizontal surface flows, TFG, and deep vertical motions.}
{The TFG were detected after oscillation filtering of intensities or Vz/Vdop flows, using a segmentation
  and labelling technique. Surface horizontal flows were derived from local correlation tracking (LCT)
and from intensities or Vz/Vdop flows. These methods were applied to Hinode observations,
2D surface results of a first simulation, and 3D Vz data of a second simulation.}
{ We find that TFG and horizontal surface flows (provided by the LCT) can be detected either from
intensities or Vz/Vdop component, for high resolution observations and numerical
simulations. We apply this method to a 3D run providing the Vz component in depth. This reveals a
close relationship between surface TFG (5 Mm mesoscale) and vertical downflows 5 Mm below the
surface. We suggest that the dynamics of TFG form larger scales (the 15-20 Mm supergranulation)
associated with 15 Mm downflowing cells below the surface.}
 {The TFG and associated surface flows seem to be essential to understanding
the formation and evolution of the network at the meso and supergranular scale.}


\maketitle

\section{Introduction}

 Knowledge  of the solar plasma and magnetic field dynamic evolution requires observation data
at all spatial and temporal scales. Likewise, for a better description of the
physical processes in the Sun, simulations need to integrate the smallest scales that are observable
on the Sun into their global
models. Currently, thanks to the capabilities of computers, these simulations are able to manage various scales
(from granulation to supergranulation and larger scales). \citet{Nelson2018} have shown
the importance of introducing simulations on scales up to the supergranulation to overcome the
convection conundrum, which is related to the introduction of high resolution in global 3D
simulations to describe correctly the Sun's observed differential rotation and cyclic dynamo action.
More particularly, \citet{Nelson2018} have imposed in their simulation small-scale convective plumes
to mimic near-surface convective downflows (plumes) from supergranular flows. Moving inward, the
plumes merge showing larger, more complex downflows. The introduction of these near-surface
plumes then produces  convective giant cells in the deeper layers through plume self-organization.

From the point of view of the observer, the determination of the flow organization on the solar surface and
deeper inside the Sun (a few megametres) is still currently a challenge. Different approaches have been
performed to improve the description of the convective-turbulent motions; these include helioseismology,
correlation tracking, morphologic techniques such as  network void or  the detection the trees
of fragmenting granules (TFG) \citep{Greer2016, Roudier2016, Berrilli2014, Rieutord2010, RLRBM03}).
Using a new helioseismic technique \citet{Greer2016} showed that supergranulation probably forms
at the surface, then  rains downward imprinting its pattern in deeper layers. The slow upflows could
be produced passively to fill the spaces between downflowing material. In that case, the upper
convection zone is driven by the surface cooling \citep{Greer2016}. 

The TFG detection is generally derived in the quiet Sun at disc centre from intensity observations
with the Solar Optical Telescope (SOT on board Hinode) \citep{Roudier2016, RLRBM03}. 
However, \citet{Malherbe2018}, using surface results of a 3D numerical simulation of the magneto-convection,
showed that TFG also form in the emergent intensity of the simulation (at $\tau=1$) and
are able to reproduce the main properties of solar TFG, such as lifetime and size, associated
horizontal motions, corks, and diffusive index, close to observations. While TFG appear to structure
the flows and magnetic field on the Sun surface \citep{Malherbe2018, Roudier2016, RLRBM03}, we do not
know the imprint of such a flow organization deeper in the Sun. In this paper, from a recent 3D numerical
simulation of the magneto-convection,  we use the vertical velocities Vz (z = 0.48 Mm to z = - 20.3 Mm)
to screen their link or not with the TFG detected on the surface. We use first  2D results
(z = 0 Mm emergent intensity and surface velocity vector) of a 24 h
sequence extracted from a longer 3D numerical simulation for testing TFG/LCT methods based on Vz detection
and for comparison with Hinode observations. A second 4 h sequence of a 3D vertical velocity component
(Vz only) is extracted from another 3D numerical simulation.

Section 2 describes the simulation data  and the solar observation obtained with SOT/Hinode satellite.
The detection of the TFG in the Vz component (equivalent to the solar radius direction) and horizontal velocity
also measured by local correlation tracking (LCT) on the Vz component are described in section 3.  That TFG and the horizontal velocity
are compared to those obtained in the emergent intensity at $\tau=1$ (section 3). The TFG identified with Hinode
data in the intensity and Doppler fields validate the detection on real solar data of same TFG in both methods
(section 4), where Vz of the simulation is replaced by the Doppler velocity.
The links between TFG found in the Vz component, in the 4 h simulation, at the surface at $\tau=1$, the Vz velocities
in depth (5, 15 Mm), and the correlation between the surface dynamics and deeper are described in section 5.
The results  are summarized and discussed in the conclusion (Section 6).

\section{Simulation data and Hinode solar observations }
 
 We used the 2014 results of the 3D magneto-convection code  (\cite{SN98,Stein2009}, review by \cite{Stein2012})
, which was not designed  to model solar TFG. This code solves the equations of mass, momentum, and internal energy in
 conservative form plus the induction equation of the magnetic field for compressible flow on a staggered mesh.
Boundaries are periodic horizontally and open at the top and bottom. Radiative heating/cooling is 
calculated by explicitly solving the radiation transfer equation in both continua and lines
assuming LTE. The number of wavelengths for which the transfer equation is solved is reduced via a
method which models the photospheric structure. Runs have dimensions 2016 x 2016 x 500 with resolution
48 km horizontally and 12-80 km vertically.  Then the horizontal field is $96Mm\times 96Mm$ and the vertical axis 
extends from the temperature minimum (0.48 Mm) down to 20.3 Mm below the visible surface.

 Two data sets of magnetohydrodynamics (MHD) simulations, described above, were used to detect the TFG. The first is a
 24 h duration sequence of the emergent intensity, surface vertical velocity (Vx,Vy,Vz), and magnetic field
 vector extracted between t = 59.6 h to 83.6 h  and a field of $96Mm\times 96Mm$. Because of the huge volume
 of the data, we restricted the analysis to the field of $48Mm\times48Mm$  ($65\arcsec \times 65\arcsec$). 
 For that sequence we have only intensity and velocity vector
 at $\tau=1$ (z=0 Mm). Hence, horizontal plasma flow (Vx, Vy) can be compared to those detected by the
LCT applied to intensity or Vz component alone.
 Because of the huge volume of the data the  pixel was rebinned from 0.065\arcsec to 0.13\arcsec and the
 time step is 60 s. In order to be in the same condition as solar observations, both quantities
 (intensity and velocity) were filtered by the Hinode point spread function (PSF) at 557.6 nm. To remove the
 5 min oscillations, we applied a subsonic Fourier filter in the $k-\omega$ space, where k and $\omega$ are the horizontal wave number and pulsation, respectively.  All  Fourier components such that  $\omega \le Cs\times k$
, where Cs=6 $km~s^{-1}$ is the sound speed and k the horizontal wave number, were retained to keep only convective motions.

 \begin{figure}
\includegraphics[width=7cm]{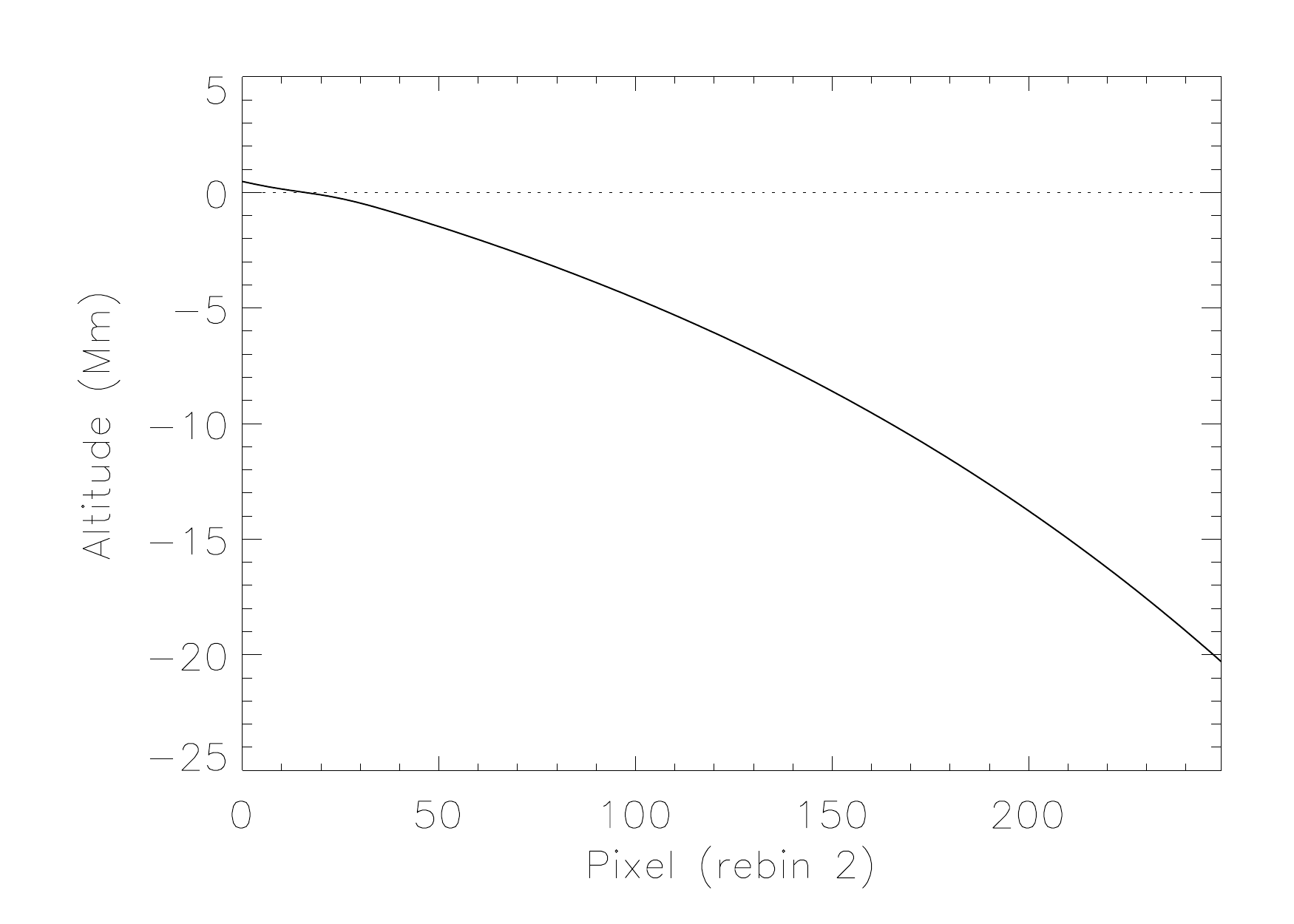}
 \caption[]{ Depth in Mm relatively to z pixels
} \label{Alti}
\end{figure}

  Because of the huge volume  of data in the second MHD simulation of 4 h duration, we used only the vertical velocity Vz as a
   function of depth between z = 0.48 Mm (top) to z = - 20.3 Mm (bottom) from t = 68.15 h to 72.25 h. The horizontal field
 is $96Mm\times 96Mm$. The other components (intensity, Vx, Vy and magnetic field) exist, but we were not able to get
 these owing to excessive data volume.
  Figure~ \ref{Alti}  shows the depth in Mm relatively to z pixels. Because of the large data volume
  the z pixel was rebinned by a factor 2, such that Vz at $\tau=1$ corresponds to the z pixel 20
  and  242 to equates to a depth of -20 Mm. Again, owing to the large amount of data, the (x,y) pixel was
  rebinned from 0.065\arcsec to 0.13\arcsec. The sequence time step is 60 s. For surface only (z = 0) the
  data were filtered by the Hinode PSF at 557.6 nm and then the 5 min oscillations were removed in the same
  way as described above for the previous simulation.

\begin{figure}
\includegraphics[width=7cm]{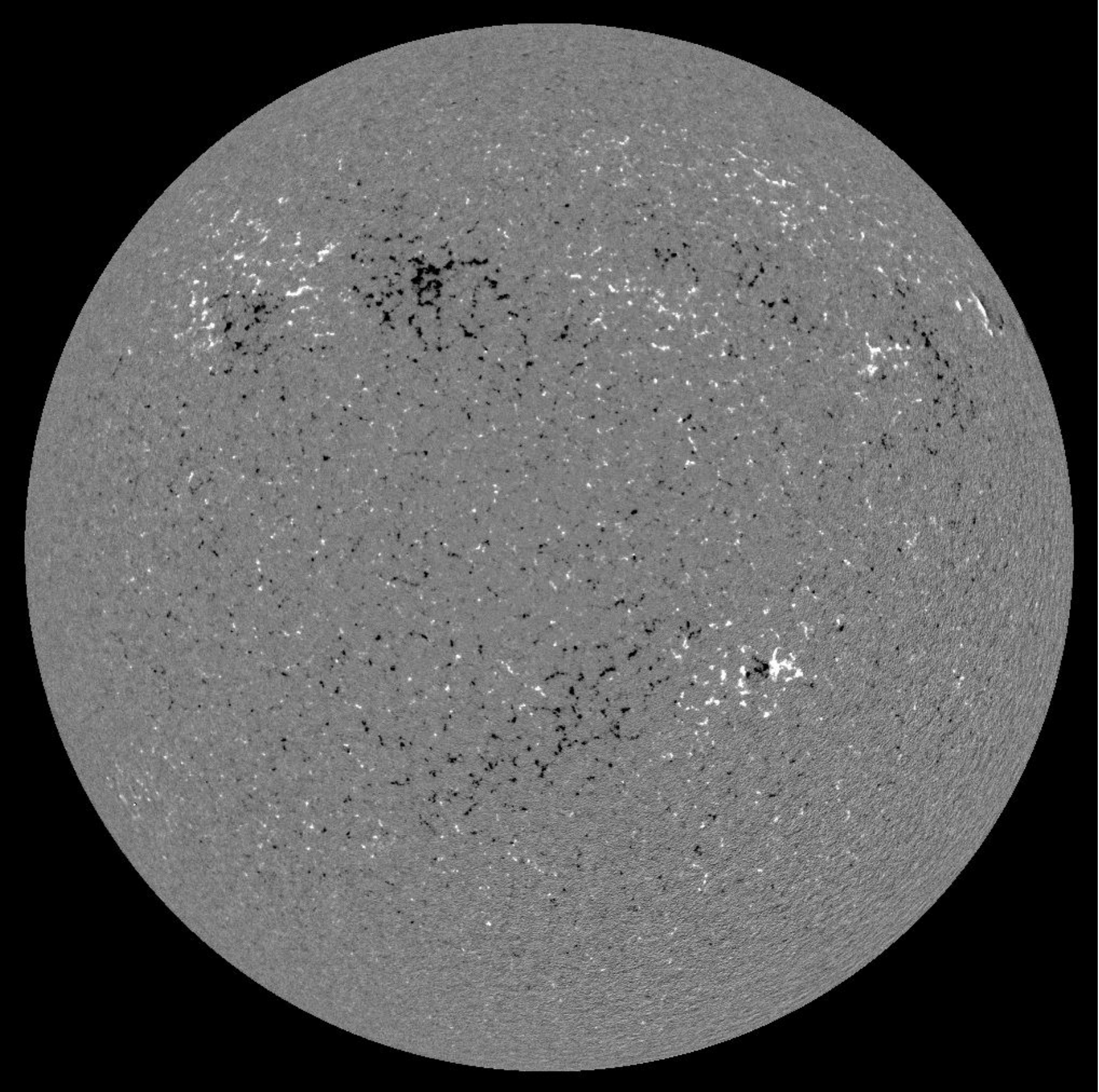}
\caption[]{ Longitudinal magnetic field context from SOHO/MDI on 14 April 2010 (11 h U.T.)
} \label{B}
\end{figure}

 In order to compare TFG detected from  solar observations in intensity and Doppler, we used 
Hinode NFI FeI 557.6 nm data, a 6 h sequence on 14 April 2010 from t = 7h to 13h U.T. 
That sequence represents 7000 spectral images with a  pixel of  0.08\arcsec rebinned to 0.16\arcsec.
The time step is 28.75s for 7 wavelengths (-12, -8, -4, 0, 4, 8, 12) pm, but the theoretical resolution
of the Lyot filter is 6 pm. Scattered light of the filter was roughly corrected. Granulation intensity
was derived from the average of wings at -12 and +12 pm at z close to the solar surface. Bright points
were  identified from core intensity at z=180km above the surface \citep{Malherbe2015}. The Doppler shifts
were computed with the bisector technique around inflexion points providing Vdop above the solar
surface at z =130 km \citep{Malherbe2015}.
Intensity and Vdop were recentred by cross correlation and filtered from 5 min oscillations.
The FeI 557.6 nm line is insensitive to magnetic fields and the bright points observed in the
intensity line correspond to hot spots that are considered as proxies of magnetic fields. No CaII H
or magnetic data are available from Hinode for this observation.  Figure~ \ref{B} gives the magnetic
context quiet Sun at the disc centre on 14 April 2010 (11h U.T.) from a MDI longitudinal magnetic field observation.
 \begin{figure*}
 \centerline{
  \includegraphics[width=18cm]{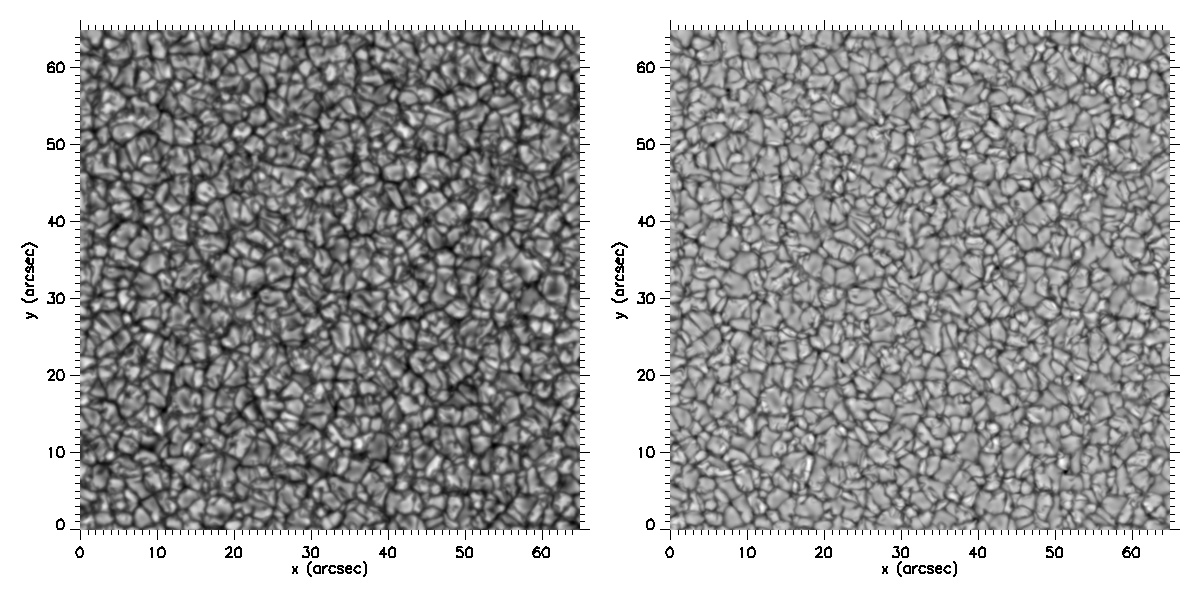}}
  \centerline{
  \includegraphics[width=18cm]{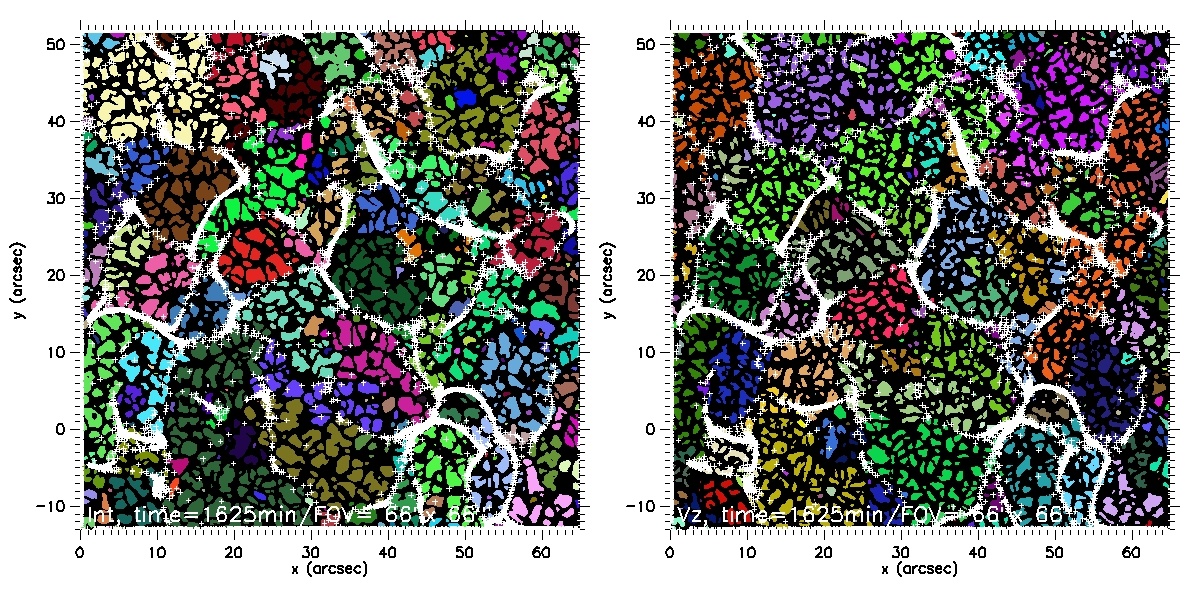}}   
  \caption{ Top : typical emergent intensity ({\it left}) and vertical velocity Vz ({\it right}) at the surface (z = 0 Mm) from the 24 h simulation. Bottom : TFG and corks (white crosses) detected from the emergent intensity at the
end of the 24 h simulation  ({\it left}); TFG detected from the velocity component Vz with corks derived from the
horizontal plasma velocity Vx, Vy ({\it right}).  The movie1.mp4  shows  the temporal evolution of the bottom figures.
  The time step is 60 s; field of view (FOV) is 65\arcsec x 65\arcsec. }
 \label{simu}
 \end{figure*}

\section{ TFG and horizontal flows detected in Vz (0 Mm) of the long sequence (24h) simulation}

\begin{figure}
\centering 
\includegraphics[width=9cm]{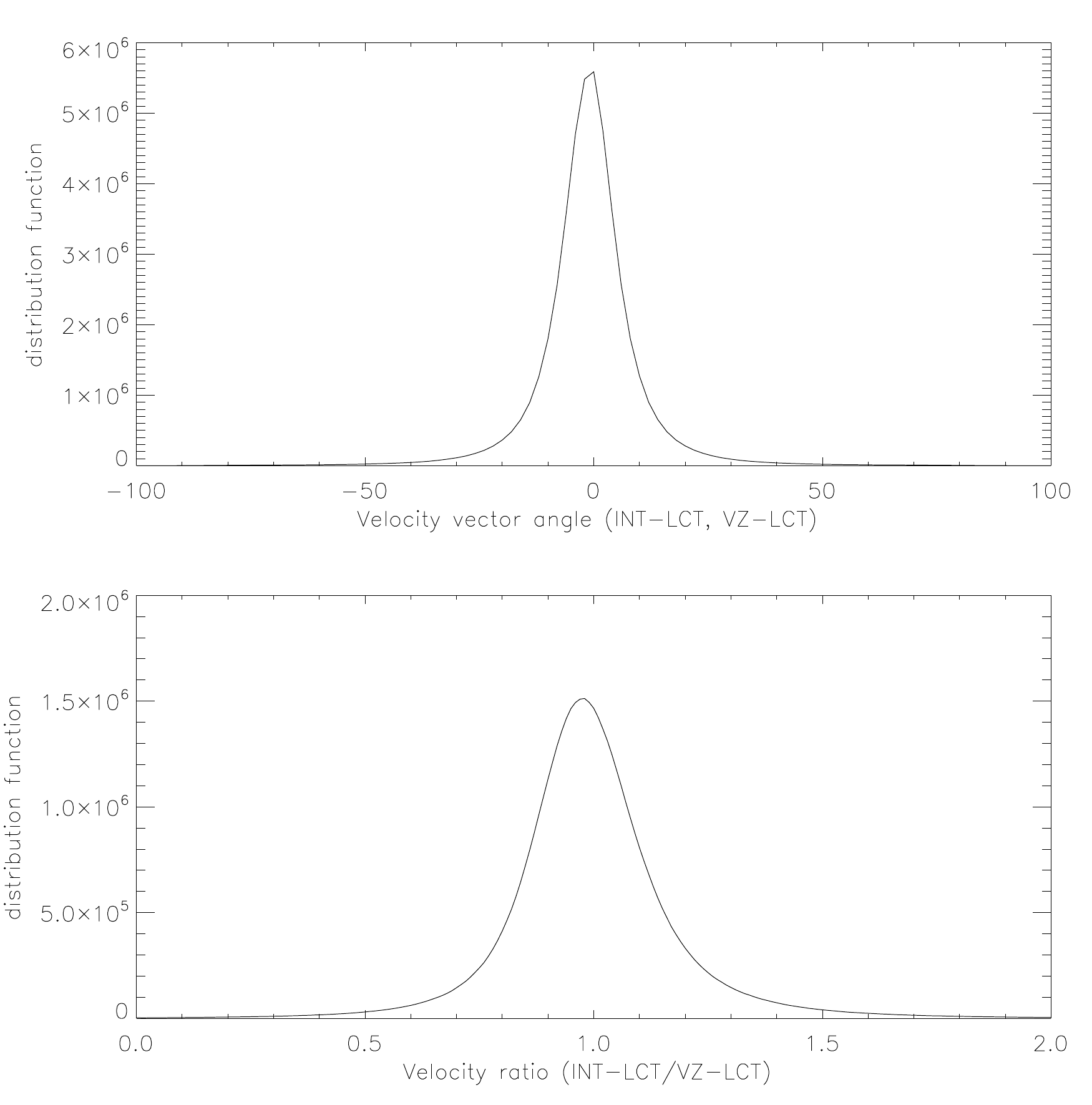}
\caption[]{ Angular gap and ratio of velocities obtained with LCT (windows 30 min. and 3\arcsec) on emergent intensity
  to Vz components.
} \label{vit}
\end{figure}

\begin{figure}
\centering 
\includegraphics[width=9cm]{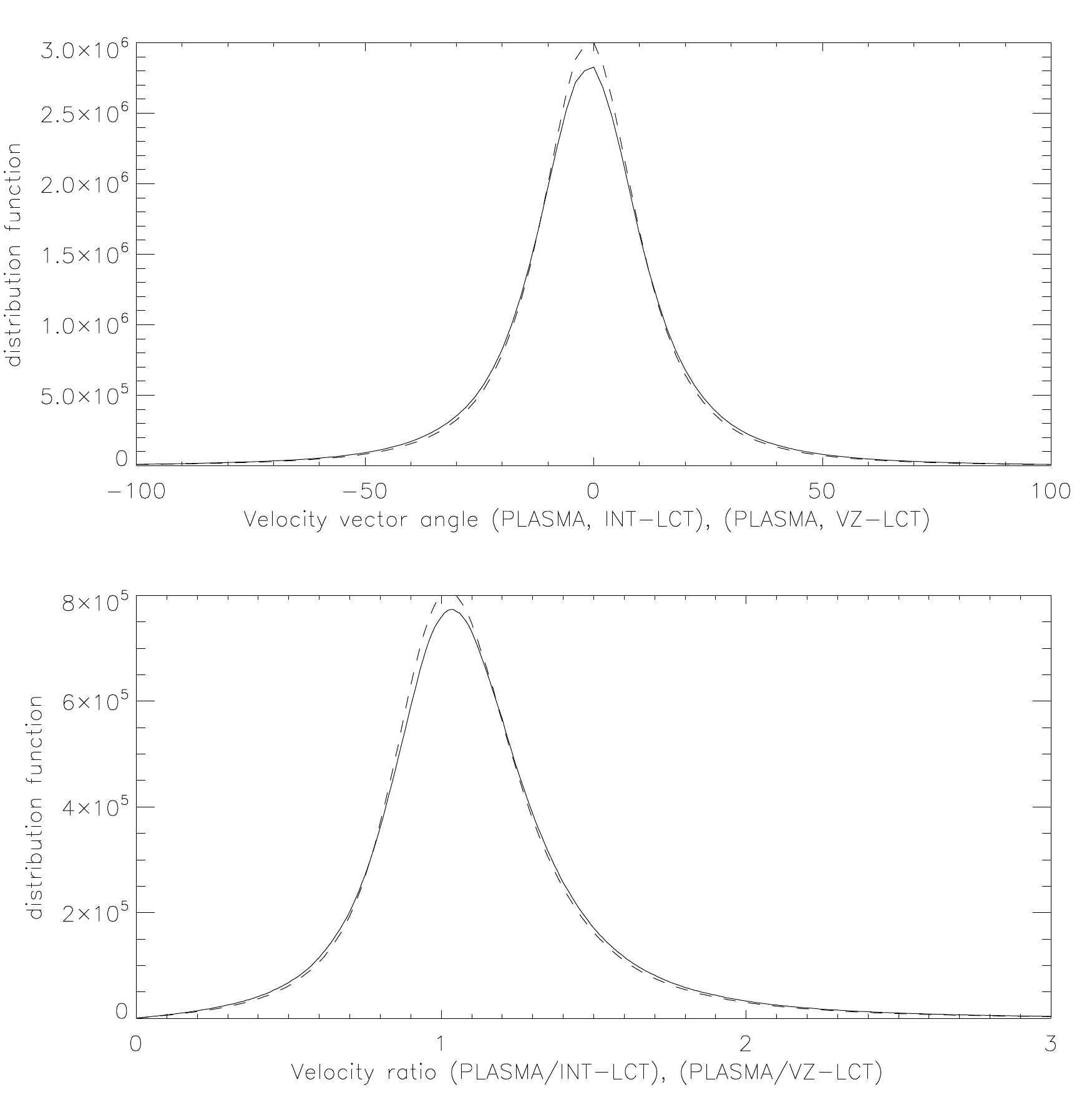}
\caption[]{Comparison between velocities of the plasma and velocities obtained with LCT
on emergent intensity (solid line) and comparison between plasma velocities and velocities obtained with LCT on
Vz (dotted line). The LCT windows 30 min and 3\arcsec 24h statistics. Top: angular shift between
velocity vectors. Bottom: ratio between velocity vector modules.
} \label{vitb}
\end{figure}

\begin{figure}
\centering 
\includegraphics[width=9cm]{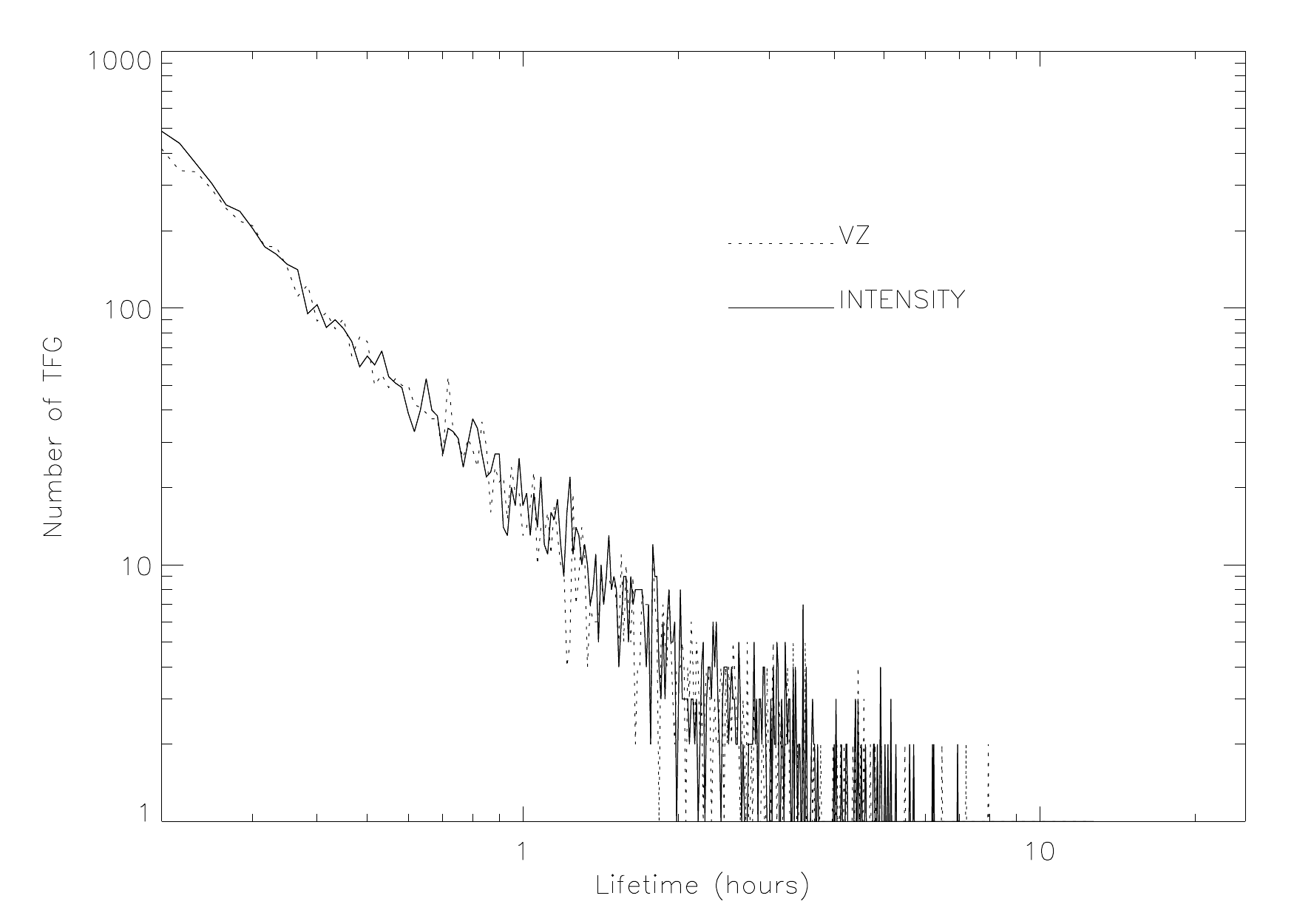}
 \caption[]{Lifetime histograms of TFG detected from  Vz ($\tau =1$ (at 0 Mm)) and emergent intensity. 
} \label{life}
\end{figure}

\begin{figure}
\centering 
\includegraphics[width=9cm]{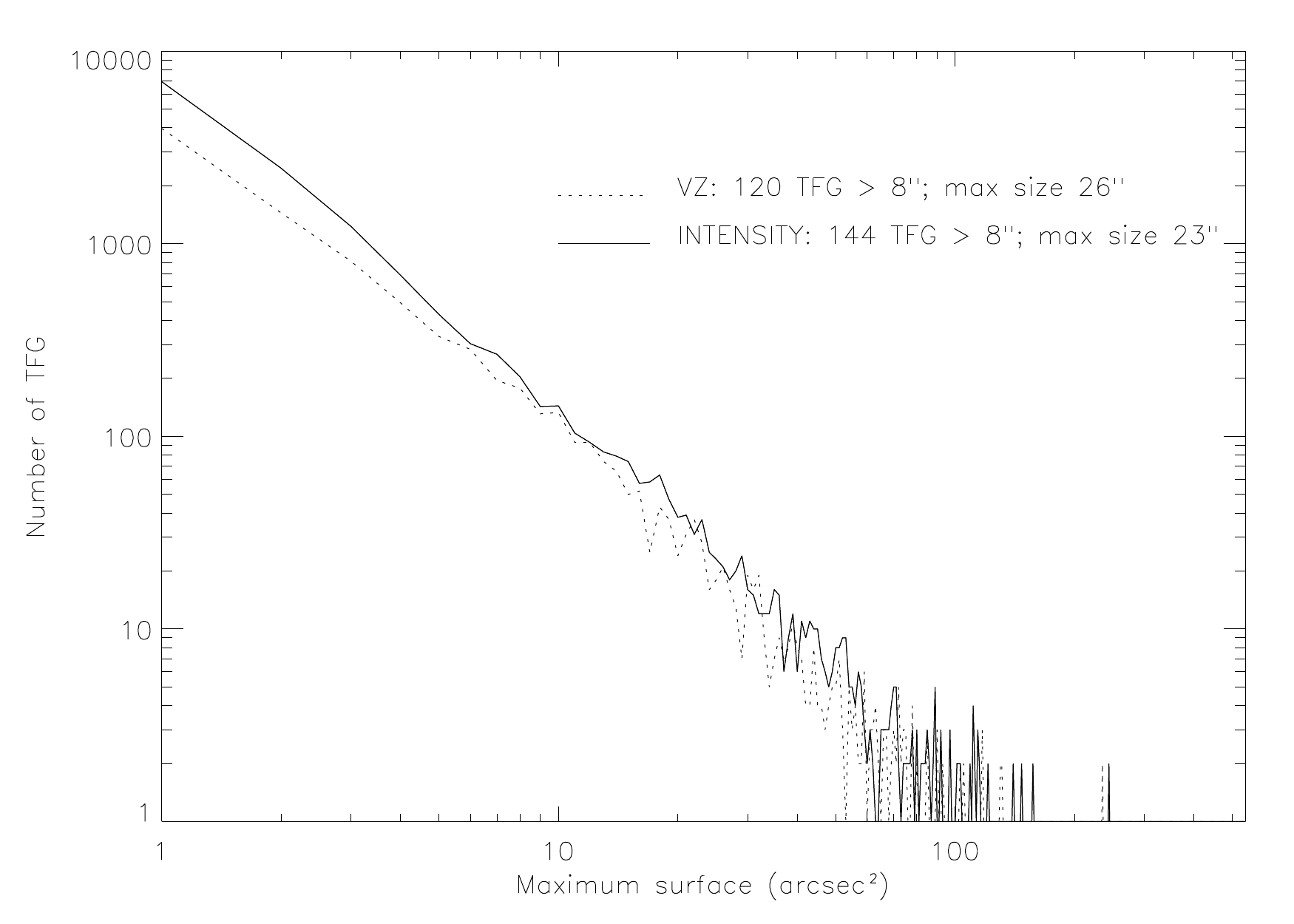}
 \caption[]{Size histograms of TFG detected from  Vz ($\tau =1$ (at 0 Mm)) and emergent intensity.
} \label{size}
\end{figure}

The first step of our analysis is to compare the TFG detected in the emergent intensity
and Vz (radial velocity) on the 24 h simulation. Both sequences  are filtered by the Hinode PSF.
The TFG are detected using a segmentation and labelling technique described in detail by \cite{RLRBM03}.
Although we can recognize each granule in grey levels at the top of Figure~\ref{simu} where intensity
and Vz are shown, the granules appear slightly different in amplitude repartition.
This is why the transformation of the intensity and Vz maps into binary maps gives a relative different
proportion of granule area to the total area of 41\% and 38\%, respectively.
The bottom of Figure~\ref{simu} shows the TFG detected from intensity or Vz segmentation at the end
of the sequence (24h) with superimposed corks (white) which move freely at the speed
of the horizontal flow (Vx, Vy) provided by the LCT (using classical 30 min and
3\arcsec windows , left) or by the plasma velocity (right). We checked that cork final positions, based on the LCT or true plasma velocity of the simulation,
are almost identical for the reasons described below.
Most of TFG can be identified in both maps (intensity and Vz).
The different temporal labelling is essentially due to the different amplitude repartition in
granules of the intensity and velocity Vz component. In both maps corks are located on the edges
of TFG and delineate supergranular scale.  The temporal evolution of the TFG and corks is shown in the animation
movie1 attached to  Figure~\ref{simu} (bottom). 
\begin{figure*}
\centering 
\includegraphics[width=18cm]{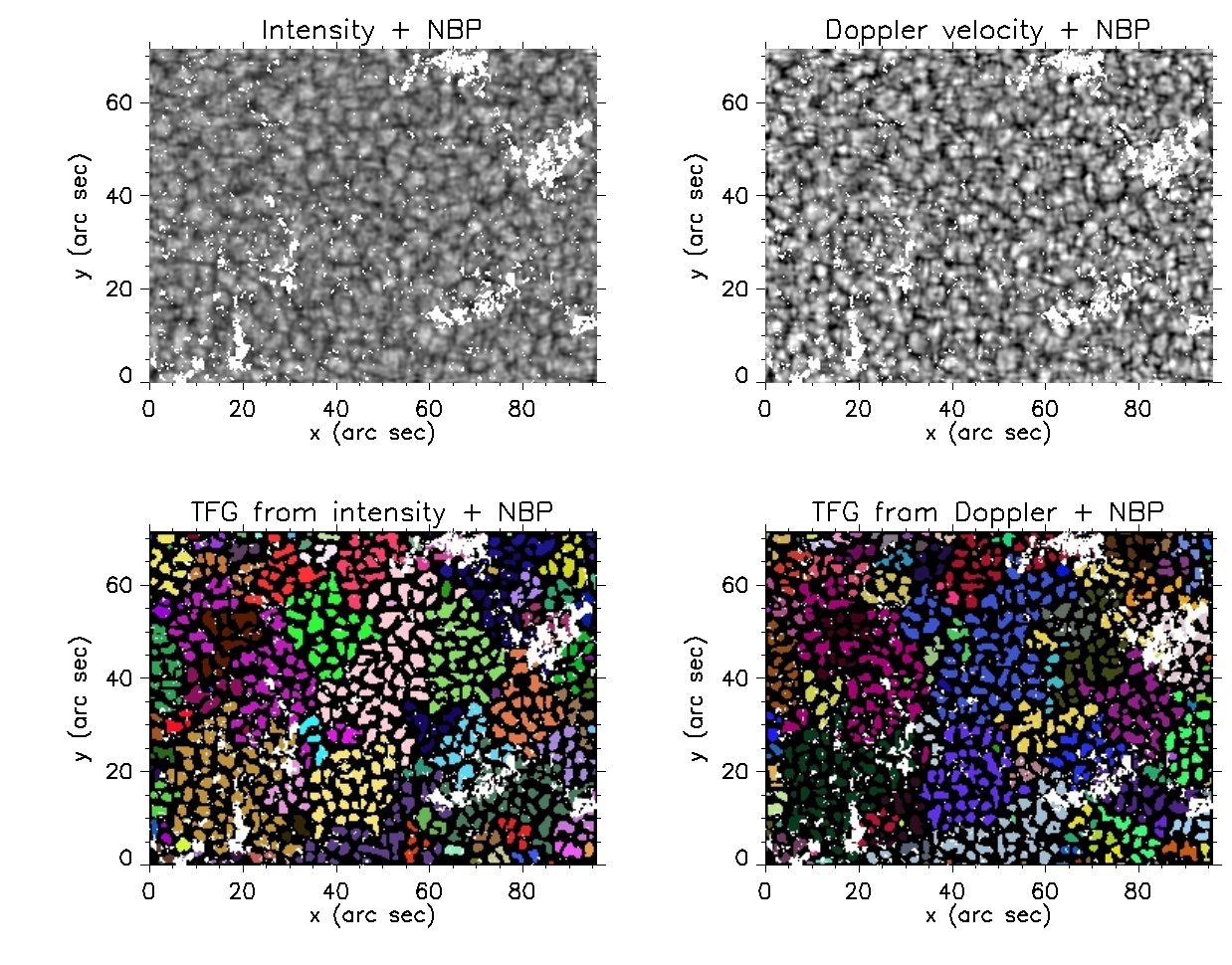}
\caption[]{ {Top:} Hinode observation with NBPs at line centre superimposed (white).
Intensity ({\it left}) and Doppler velocity ({\it right}).
Bottom: TFG with NBPs superimposed (white).  TFG derived from intensities ({\it left}) and TFG derived
from Dopplershifts ({\it right}).
} \label{TFGintdophin}
\end{figure*}

\begin{figure*}
\centering 
\includegraphics[width=18cm]{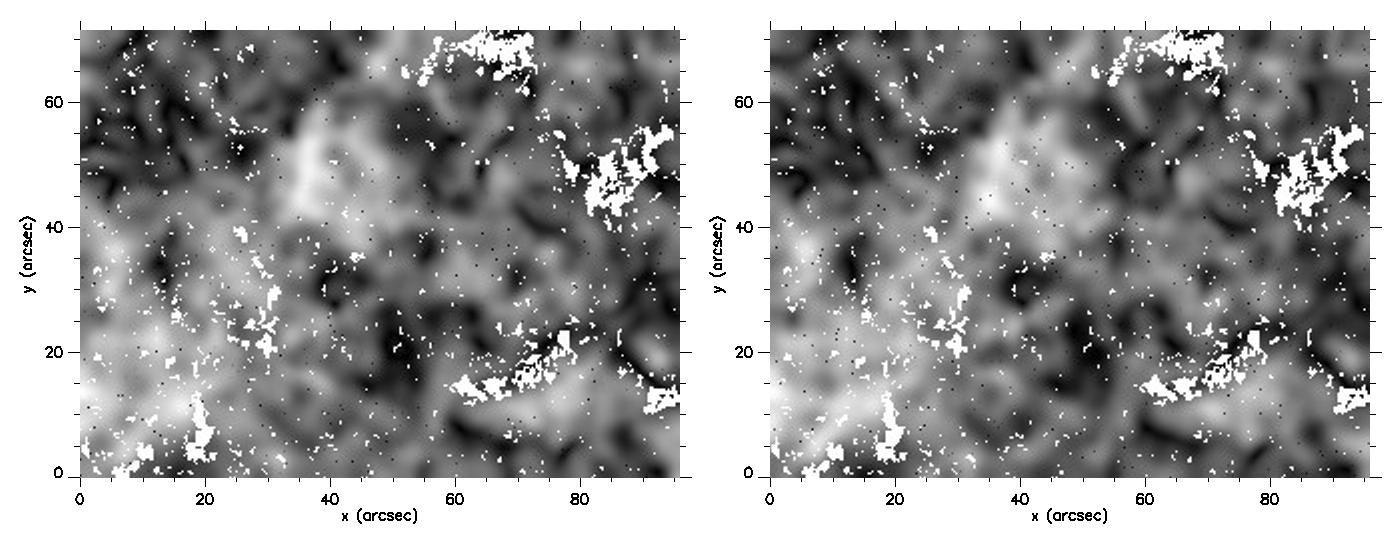}
\caption[]{ Six hour averaged module of the horizontal velocities Vh computed with LCT (windows 30 min, 3\arcsec)
  on intensity ({\it left}) and Doppler velocity ({\it right}) with NBPs (line centre) superimposed (white).
} \label{Vhhin}
\end{figure*}

\begin{figure*}
\centering 
\includegraphics[width=18cm]{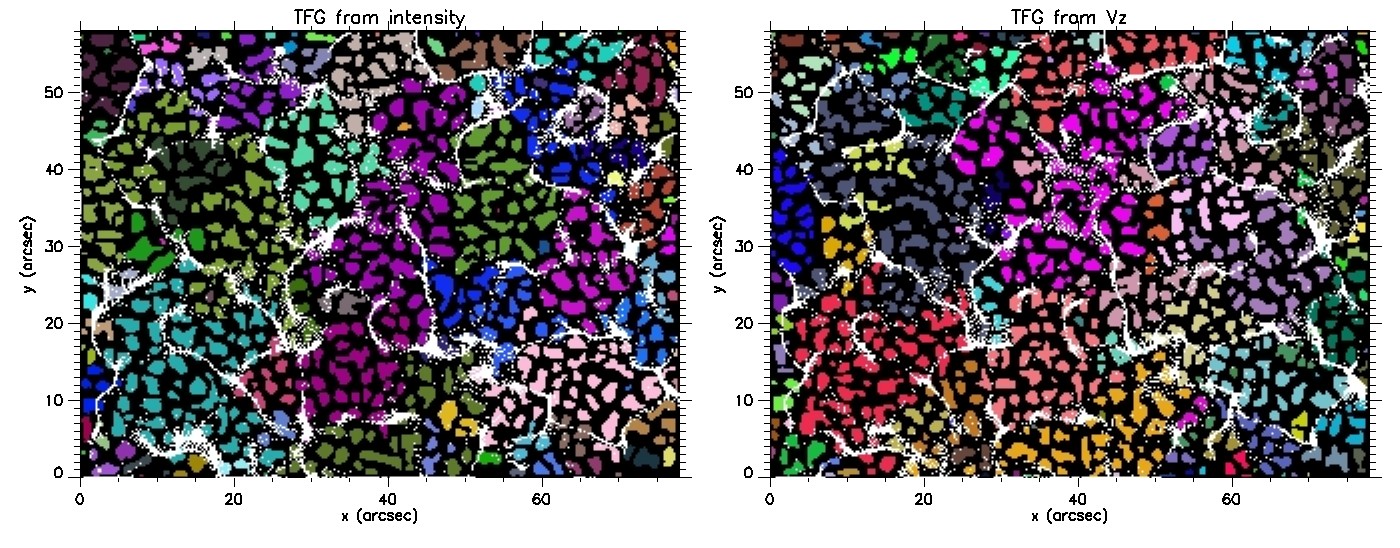}
\caption[]{TFG detected from the intensity and Doppler field  with corks superimposed (in white).
} \label{TFGcorkintdop}
\end{figure*}

The correlation between the velocity obtained via the LCT, on
both fields (intensity and Vz) gives a coefficient of 98\%.  Figure~\ref{vit} shows the angular gap and ratio
repartition between the two LCT velocity measurements giving the maximum error on the angle and module determination:
$\pm5^\circ $ in 50\% and $\pm16^\circ$ in 90\%  of the cases, and 9\% in 50\% and 29\% in 90\% of the cases on the module.
These values indicate a very good correspondence between
the horizontal velocities measured with LCT on intensity and Vz fields. Figure~\ref{vitb} exhibits the
comparison of the simulation plasma horizontal velocity  with velocity measured by LCT on the intensity and Vz fields.
In that case the distributions  are larger due to, in large part, the spatial and temporal averaging windows
used in the LCT velocity determination.

Figure~\ref{life} shows the similarity of the histogram of the TFG lifetimes for emergent intensity and
velocity Vz. That  distribution  function of  family  lifetimes is typically a power law which is in good
agreement with previous results \citep{Malherbe2018}. The comparison in Figure~\ref{size} of the histograms of TFG maximum area
during their lifetimes exhibits a small difference in the area that is smaller in size owing to
the lower proportion of granule area in the Vz component; however, for the larger areas the behaviour is very
similar.

\section{Hinode TFG and horizontal flows detected in Doppler and intensity}

 Our goal is to detect TFG on emergent intensity and Doppler velocities on real solar data to compare  to the previous results obtained with the simulation,  where now the  Vz of the simulation
 is replaced by the Doppler velocity (Vdop). From spectral images observed by Hinode, we can build temporal
 sequences in intensity and Dopplergram. The Doppler is deduced from the observed line profile while the Vz
 plasma velocity is computed for each altitude of the simulation. So the Dopplergrams reflect, at the disc
 centre, the radial velocities in a range of altitudes with a maximum of the contribution function
 around 130 km \citep{Malherbe2015, Sheminova1998}.
 Figure~\ref{TFGintdophin}  (top) shows intensity and Dopplergram at the end of the sequence and reveals
 the difference of appearances of both fields. This reflects the different altitudes of formation
 (intensity in line wings, dopplershifts at inflexion points), integration along the line of sight, which is also
 degraded by stray light and limited spectral resolution of the NFI filter.
 However, the solar granules are identifiable in both fields.
 As for the simulation sequences, the transformation of the intensity and Vdop maps into binary maps
gives a different relative proportion of granule area to the total area of 38\% and 37\%, respectively.
Figure~ \ref{TFGintdophin} (bottom) shows the TFG  detected for both components (intensity, Doppler) at the end
of the sequence (6 h) and can be identified in both maps. The different temporal labelling is essentially
from the different amplitude repartition in granules linked to the altitude of formation and integration.

Figure~\ref{Vhhin} shows the 6 h averaged module of the horizontal velocities Vh derived from LCT
(classical windows 30min, 3\arcsec) applied on intensity and VDop fields.
The Vh fields measured on intensity and Doppler maps are
similar in amplitude and repartition. The bright points detected in the central part of the line  557.6 nm used as proxy of the longitudinal magnetic component are located in lowest amplitude of the horizontal velocities
forming aligned structures which delineate a supergranule in the central part of the figures.
Figure~\ref{TFGcorkintdop} shows the TFG detected from the intensity and Vdop with final positions of corks
superimposed (white crosses). Corks delineate the TFG at the meso or supergranular scale and are located close
to the proxies of the magnetic field, network bright points (NBPs) (see figure~\ref{TFGintdophin}).
The spatial correspondence between corks and the proxy of the magnetic field indicates a good determination
of the flow field also related to the TFG evolution.

\section{ Surface TFG and deep downdraft predicted by the (4h) 3D simulation  of vertical velocity Vz  }

\begin{figure}
\centering 
\includegraphics[width=9cm]{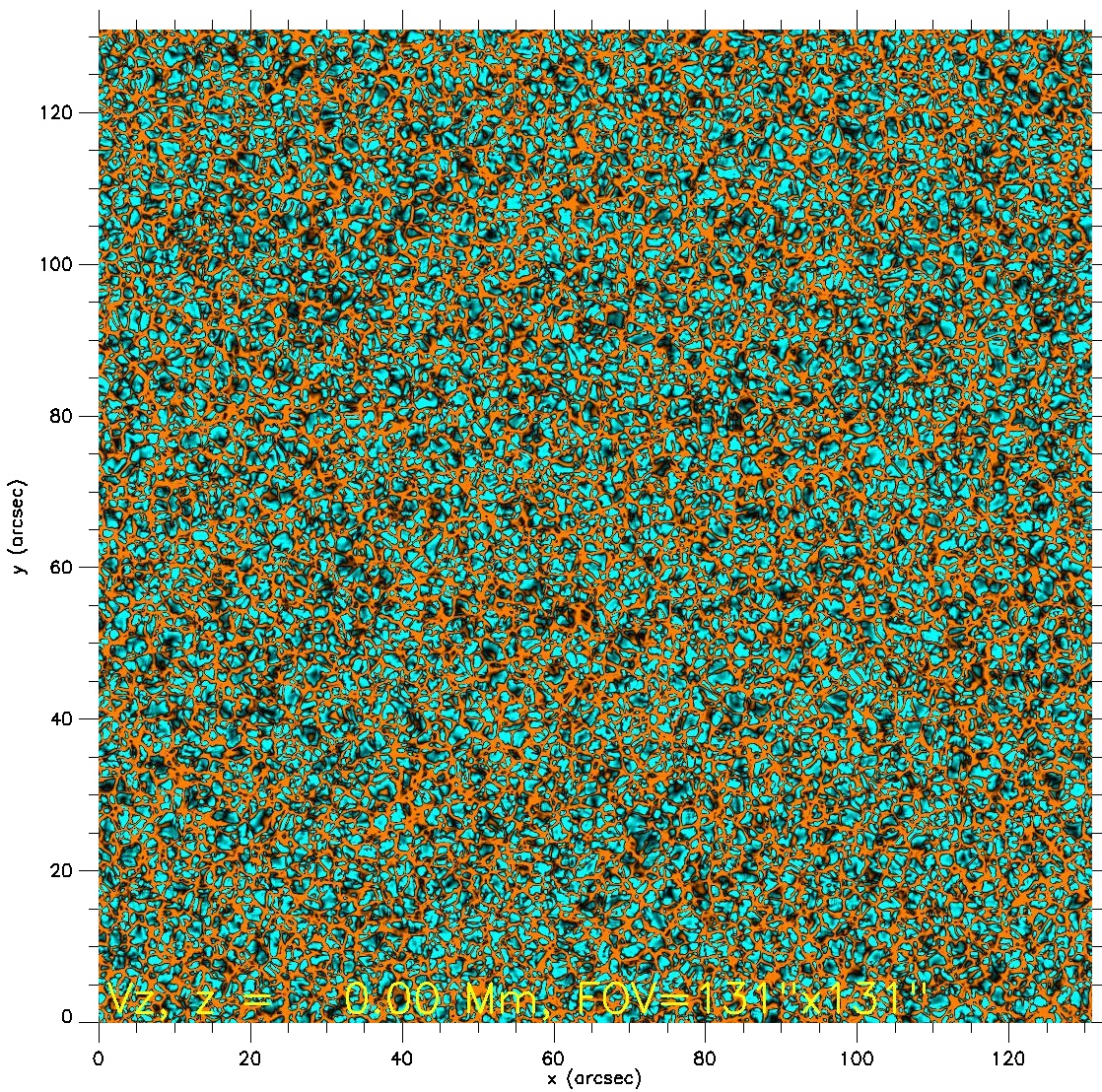}
\caption[]{  Vz  of the simulation at the altitude z= 0 Mm. Green corresponds to the granule (upflows) and
  orange to intergranule (downflows). The movie2.mp4 shows the vertical velocity (orange/green = downward/upward) of
  the simulation at different depths from +0.5 Mm (above the surface) to -20 Mm (below the surface). FOV is 131\arcsec x 131\arcsec.}
\label{vz32}
\end{figure}

\begin{figure}
\centering 
\includegraphics[width=9cm]{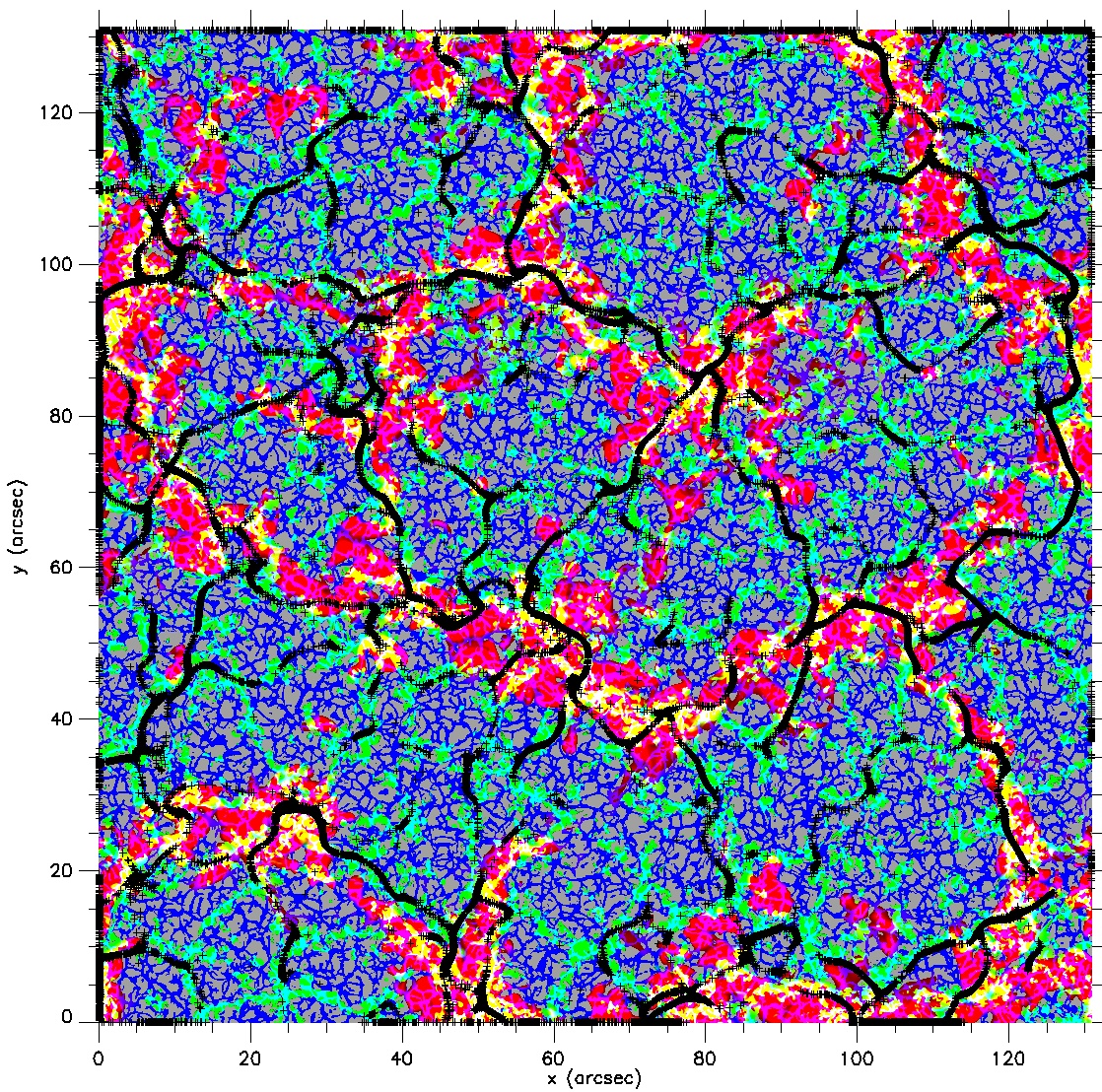}
\caption[]{Vz downflows visible at different depths of the simulation with superimposed corks (in black).
  Blue corresponds to z= 0 Mm (intergranules), green to z=-5 Mm (structures with TFG scale) at the same
  time ,and red to z=-15 Mm where the supergranular scale (matching about 10 TFG) is observed.
    The movie3.mp4  shows the  temporal evolution of that figure. Time step is 60 s; FOV 131\arcsec x 131\arcsec.}
\label{vzz}
\end{figure}

The 4 h simulation of the vertical velocity Vz (Figure~ \ref{vz32}) allows us, for the first time, to study the link between surface properties and downflows from depth between z = 0.48 Mm (top) to z = - 20.3 Mm (bottom).
 The vertical velocity Vz from the top to the bottom is shown in the animation movie2 attached to Figure~ \ref{vz32}.
  The downflows at different depths (z= 0, -5, -15 Mm) with superimposed corks in black (evolution described below) are shown in
  the animation movie3 attached to Figure~ \ref{vzz}. That movies immediately reveal the existence of three downflowing scales: intergranules (blue), the mesoscale (green) and the supergranulation scale (red).
At the bottom of our data cube, the larger scales are visible because downflows are
 collected and mixed to form a larger structure with depth.

\subsection{ Corks and downdraft location with depth}

\begin{figure*}
\centering 
\includegraphics[width=18cm]{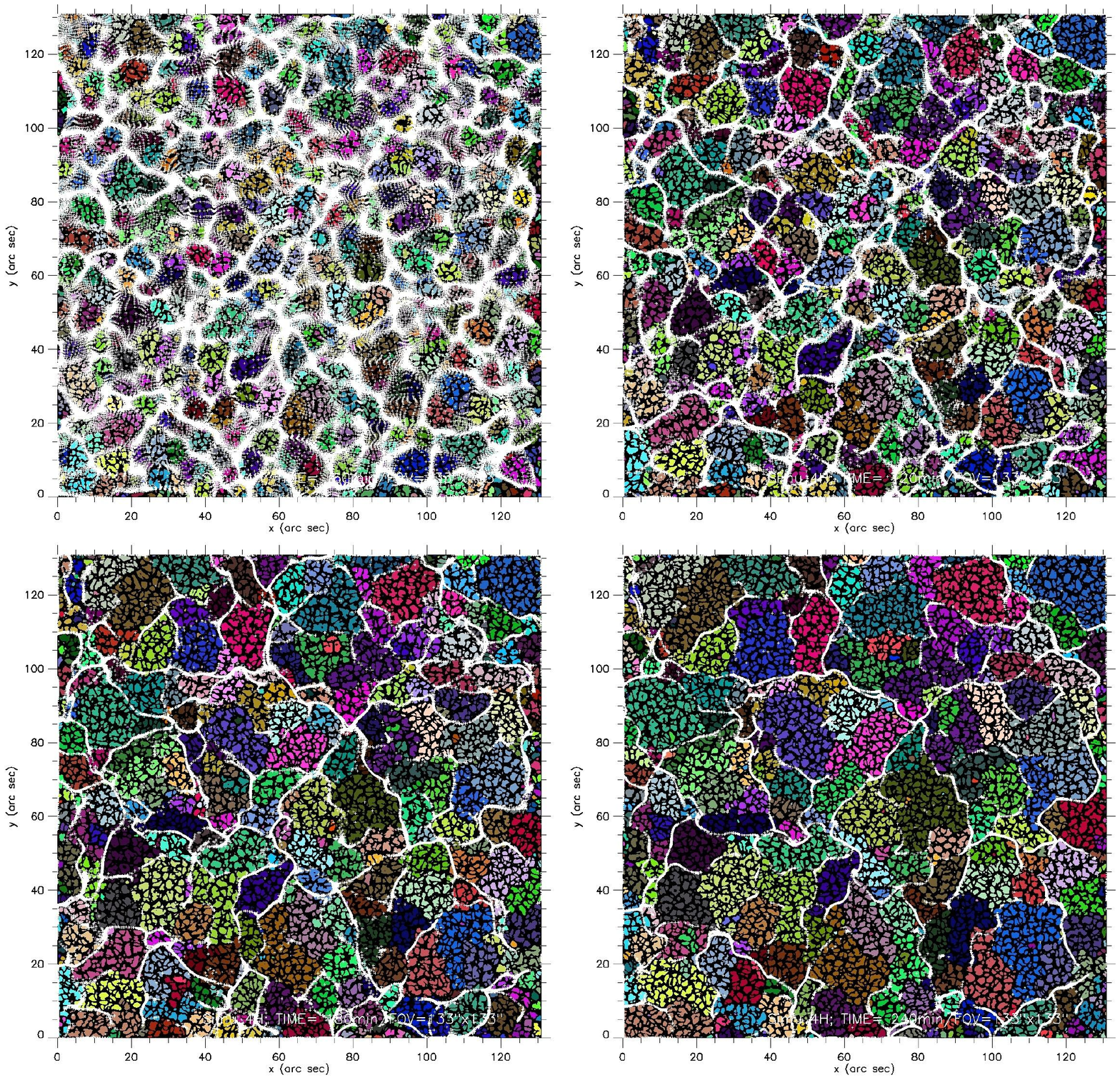}
\caption[]{ Detected TFG on the 4 h Vz simulation at z = 0 Mm with corks superimposed (white) at different
  times as follows: 60, 80 ({\it top}), 120, and 240 ({\it bottom}) minutes.
} \label{TFGcorks}
\end{figure*}

\begin{figure*}
\centering 
\includegraphics[width=18cm]{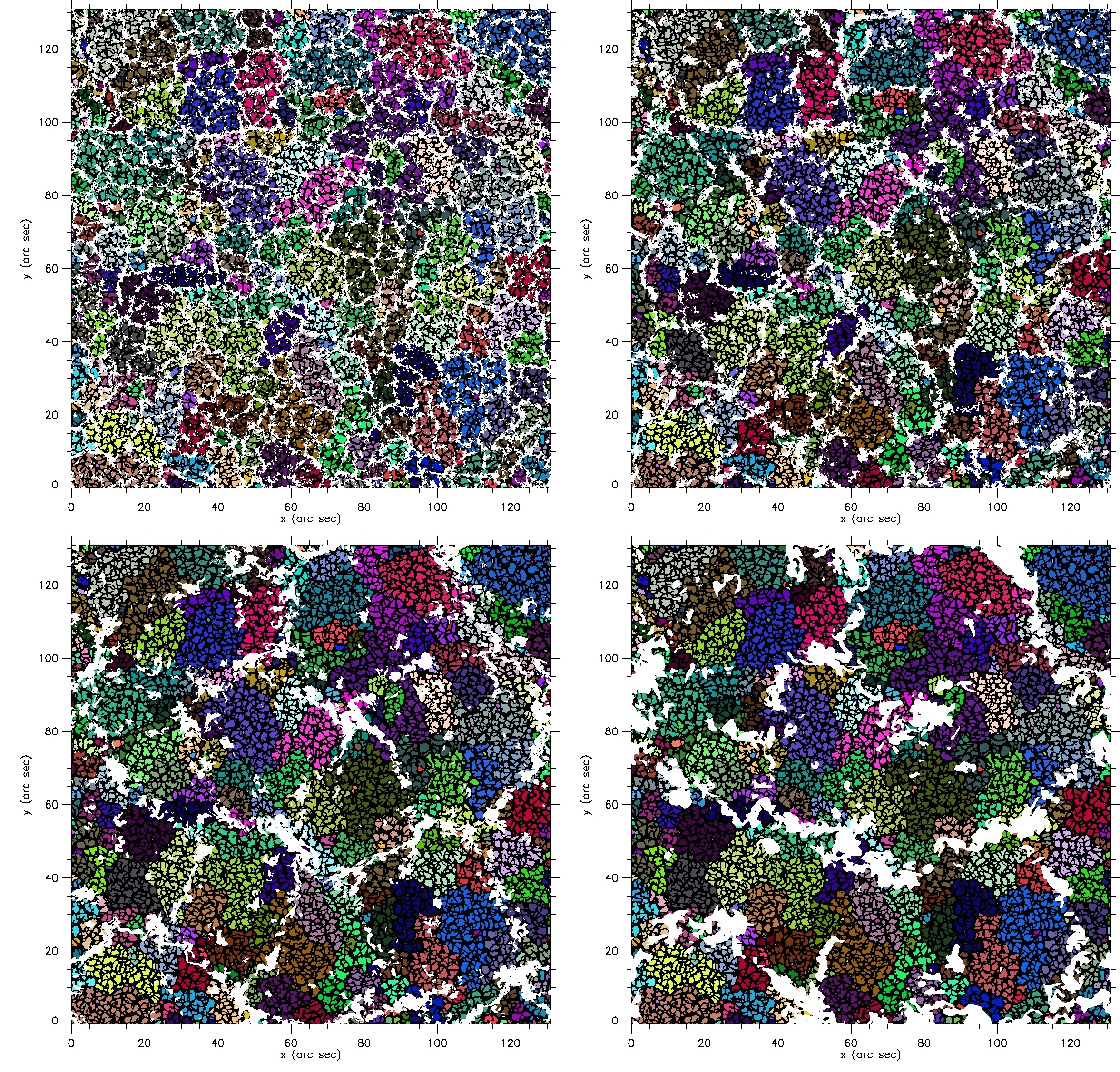}
\caption[]{ TFG detected from the Vz component at z = 0 Mm with superimposed downflows (white) at different
  depths below the surface as follows: 3.0 Mm ({\it top left}), 5.0 Mm ({\it top right}), 8.0 Mm ({\it bottom left}), and
  15 Mm ({\it bottom right}). Movie4.mp4, movie5.mp4, movie6.mp4, and  movie7.mp4  show the evolution of the families at the surface (z=0) and the downflow (white) at different depths of 3.0, 5.0, 8.0, and 15.0 Mm
    below the surface, respectively. Time step is 60 s; FOV  131\arcsec x 131\arcsec.
} \label{TFGDOPall}
\end{figure*}

The horizontal velocities obtained with LCT (windows 30 min and 3\arcsec) on the Vz component at the surface
(0 Mm) allows us to compute the corks trajectories during the 4 h sequence.  At the surface (z=0 Mm) the corks
diffused by the horizontal flows are located between TFG (Figure~ \ref{TFGcorks}) as previously
observed \citep{Roudier2016}. At the end of the sequence, corks form a larger scale than TFG ,which can
be compared to the downflowing ribbons at 8 or 15 Mm. (see for comparison figure~\ref{TFGDOPall}).
This indicates the intimate link between the horizontal flow at the surface and the downflow network
in the deeper layers.

 \subsection{ TFG and downdraft location }

  First, we tried to detect  correspondence between the location of large amplitude downdrafts visible at 0 Mm
  and at a depth of 20 Mm. The vertical location of the  downdraft at 20 Mm depth corresponds  in half
  of the case upflows at 0 Mm  and vice versa between 0 Mm and 20 Mm. This indicates clearly  that larger amplitude downdrafts
  observed at the surface do not reflect the downflow location in depth. This probably explains why heliosismology does not
  find coherence beyond 7 Mm and has difficulty detecting supergranules in depth. The second approach is to locate downflows
  at different depths relative to the TFG that structure flows at the surface (0 Mm).

   The analysis of sequence (4 h) of the  vertical velocity Vz as a function of depth between
  z = 0.48 Mm (top) to z = - 20.3 Mm  allows us to locate the TFG relative to the downflows.
  Figure~\ref{TFGDOPall} shows the TFG detected during 4 h  at 0 Mm depth with superposed downflows
  at different depths of 3.0, 5.0, 8.0,  and15 Mm.  The temporal evolutions at those different depths are shown in the
    animations movies 4, 5, 6, and 7  attached to Figure~\ref{TFGDOPall} .
  In the first depth of 3 Mm, no special link between TFG
  and downflows at the depth are visible. At 5 Mm depth, we clearly observe the downflows at the limit
  of the TFG indicating a connection of that flow and the granule evolution at 0 Mm.
  For deeper depth (8.0, 15.0 Mm) the downflows are still between TFG but include several of them.
  This is due to the limited duration of our sequence (4 h) where the TFG are not fully developed in size.
  We observe some of these as older branches of larger TFG. The spatial coherence due to families of granules at
  the surface sweeps the strongest movements, high horizontal velocities inside the TFG cells at z=0 Mm, towards
  their boundaries where almost null horizontal velocity at z=0 Mm corresponds to deep vertical flows 
  \citep{Roudier2016}. This generates continuous descending motion, at meso and supergranular scales,  which penetrate more deeply up to 20 Mm.

  \subsection{ Module of the horizontal velocity and downdraft location }

 \begin{figure*}
\centering 
\includegraphics[width=18cm]{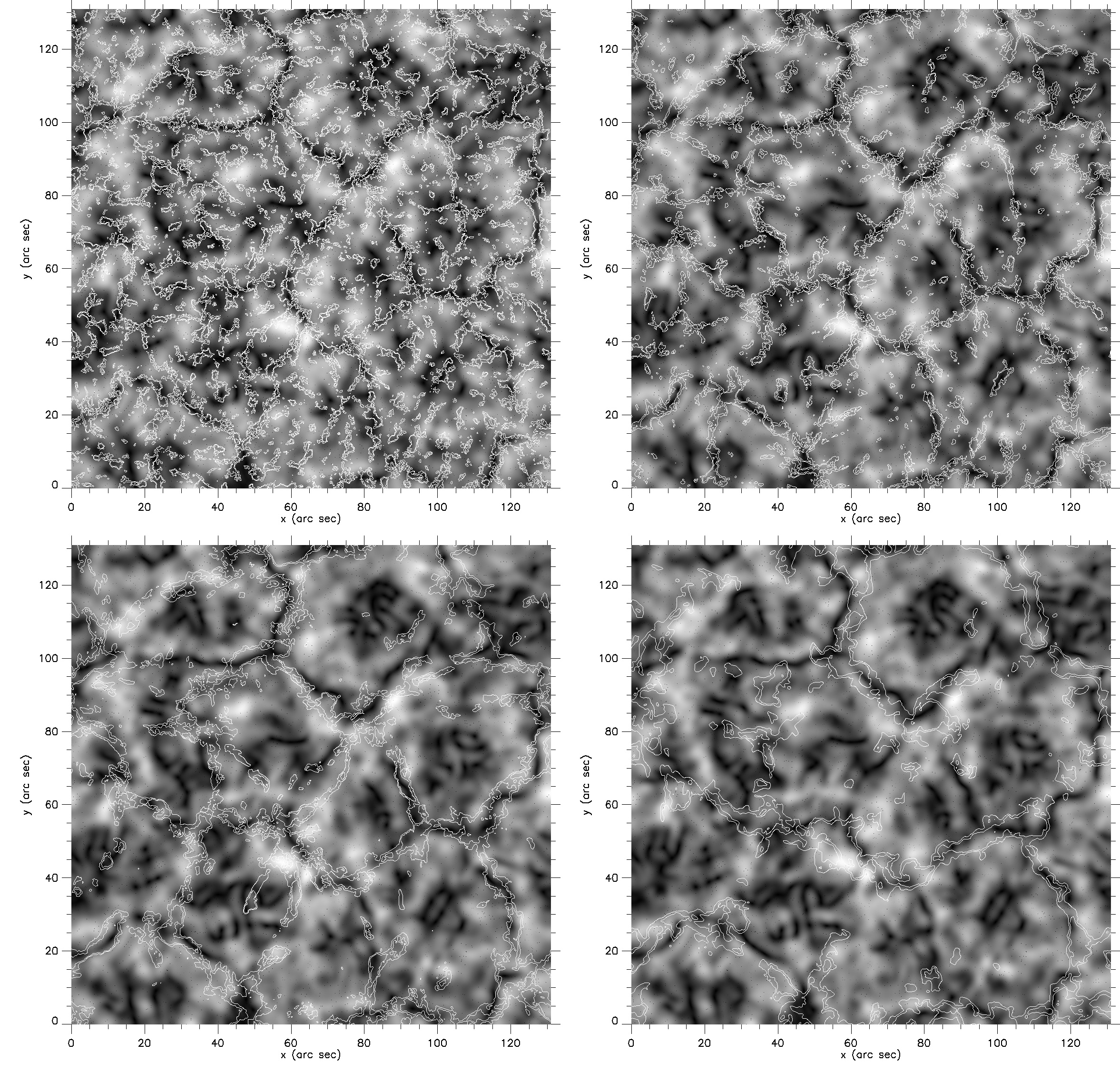}
\caption[]{ Four hour averaged horizontal velocity (Vh) module at z = 0 Mm with superimposed downflows
(white contours) at different depths below the surface: 3.0 Mm ({\it top left}), 5.0 Mm ({\it top right}),
8.0 Mm ({\it bottom left}), and 15 Mm ({\it bottom right}).
} \label{modvhDOPall}
 \end{figure*}

 Figure~\ref{modvhDOPall} shows in greyscale the 4 nh averaged horizontal velocity module (Vh)
 computed with LCT (using classical windows of 30 min and 3\arcsec) at 0 Mm on Vz component, with superposed
 downflows (white line) at different depths of 3.0, 5.0, 8.0,  and15 Mm.
 Almost null horizontal velocity regions (black) at z = 0 form ribbons corresponding to the large-scale downflows observed
 from 5 to 15 Mm. That also corresponds to the limits of the TFG (see figure \ref{TFGDOPall}) and indicates the link between TFG evolution at the surface and downdraft flows in the deeper layers  of the 20 Mm box.
 In addition, the maxima of Vh amplitude (inside TFG but near boundaries) are observed
close to the low Vh amplitude ribbons and also to the TFG borders \citep{Roudier2016}.
 If we now look at the dynamics of the downflows at 3 Mm depth, the second
 part of the movies (movie3 and movie4 ) reveals the proper motions of the downflows inside TFG to
 the limit of the TFG.
 Thereby the downflows are collected at the boundaries of the TFG, which also merge  during their evolution. This process seems
 to generate structures on a larger scale which are linked to the downflows observed in depth of the simulation. This can be compared
 with the scenario described by  \cite{Greer2016} in which the supergranulation appears to form at
 the surface and rains downward, imprinting its pattern in deeper layers. The analysis of our short simulation sequence (4 h)
 adds to this view the evolution of the TGFs (expansion, proper motions, mixing),  which generates horizontal flows at the surface
 (0 Mm) whose action sweeps downflows at their borders.
 Thus  we can draw a scenario where the collective effect of explosive granules, which form TFG, tend to form
 at their limits continuous descending motions that aggregate deeper on a larger scale downflow.
The imprint on the surface motions can be observed at the border of TFG in the form of weak horizontal velocity
amplitude but inside TFG, higher (and diverging) velocities are present. Table 1 summarizes the quantitative mean values (Velocities and divergence) measured on different locations and from corks motions.

\section{Conclusions}

The main goal of our analysis was to demonstrate that TFG and horizontal velocities issued from LCT
can be detected either in intensity or vertical velocity (Vz) field sequences, both in simulations or
in disc centre observations (in that case, Doppler shifts replace the vertical velocity component).
We checked that TFG and dynamics exhibit in surface results of a first simulation the
same properties as those observed with Hinode. Using data of a second simulation providing
Vz in depth, we studied TFG formation and evolution in relation with vertical flows below the surface,
in unobservable layers.

From Hinode/NFI observations (disc centre), we found that TFG are detectable either in intensity or
Doppler shifts and are very similar with both methods. As in previous works \citep{Roudier2016,RRBRM2009,Roudier2004} the magnetic field (in this case a proxy from NBPs) is located at the border of TFG. Horizontal velocity fields (from LCT) also play  an important role because the slowest flows match the boudaries of TFG and form a supergranular
scale.

The TFG detected in intensity and Vz velocity, in the surface results of the 24h simulation,
are also similar, indicating that both methods can be used. This is also the case for horizontal
flows derived from LCT applied either to intensity or Vz, which are highly correlated.

The analysis of the 4 h simulation of the vertical velocity Vz as a function of depth between z = 0.48 Mm
(top) to z = -20.3 Mm (bottom) reveals the intimate relation between the TFG detected at
the surface at z = 0 (on the Vz component) and deeper downflows.

In the first megametres depth, we observe no special link between TFG and downflows. At 5 Mm depth, downflows
are clearly located below the boundaries of the TFG. In deeper layers (8.0 and 15.0 Mm) the downflows are still
between TFG but include several of them. This is because of the limited duration of our 4 h sequence where
TFG are not fully developed in size. Old branches of larger or new developing TFG are observed. Our
analysis shows that the spatial coherence observed at the surface and the horizontal and vertical velocities is
from families of granules which form TFG. The collective effect of granules creates  a large
divergence inside TFG that allows concentration of downflows at TFG frontiers. Through TFG evolution, this
process leads to a supergranular scale build-up. We observe that TFG sweep out the strongest movements towards
their boundaries, thus generating continuous descending flows on aligned structures, which penetrate more deeply
down to z = -15 Mm.  The TFG and associated surface flows seem to be essential to understanding
  the formation and evolution of the network at the meso and supergranular scale. The role of the TFG relative to the
  formation and evolution of the supergranulation appears important. First, \cite{Roudier2016} showed that the maxima of
  the horizontal velocity module is  intimately  related  to  the  life  of  TFG  through  their location, strength,
  and birth date.  The frequent occurrence of horizontal velocity module patches and the interaction between
  families produces several events that contribute to the diffusion of the magnetic field and the photospheric network
  in the quiet Sun \citep{Roudier2016}. Second, in the present study, we observe that TFG collect downward motions
  at their borders (where horizontal flows vanish) which are also detected in the deeper layers. These observations 
  seem  to indicate the  crucial role of the TFG in the dynamic of the surface turbulent convection and formation of the
  quiet network and in the deeper layers (few megametres).

  It is challenging today to conclude that TFG could be one of the drivers of the  supergranulation and the magnetic
  field diffusion in the quiet Sun but we know now the importance of the TFG in that region of the Sun. To test the real
  role of the TFG, we have to analyse longer temporal  series to observe if the boundaries of super granulation cells are
  formed on the surface.

 \begin{table*}
   \centering
 \caption{  Quantitative mean values measured on different locations  at z=0Mm and from corks motions.}
   
\begin{tabular}{|c  c  c  c  c c c|}         
\hline                       
QUANTITIES: & horizontal  & vertical  & vertical  & vertical & divergence*1000 & REMARKABLE  \\
                &   velocity &   velocity z=0 &  velocity z=-5 & velocity z=-15&  & PROPERTIES \\
                &  in km/s  &     in km/s &   in km/s & in km/s&  in $s^{-1}$ &   \\
\hline
CORKS         &      0.34   &    -0.38   &     -0.60     &    -0.17     &     -0.65 (converging)&  (1) \\
\hline 
NEAR TFG BORDERS   &   1.33     &     0.10    &       0.03    &       -0.05    &         0.03   &  (2) \\
\hline
TFG CENTRE     &     0.26    &   -0.07     &    0.18    &      0.10  &         0.20 (diverging) &  (3) \\
\hline
\end{tabular}
         
\label{table:1}      
\end{table*}

  \begin{acknowledgements}
    This work was granted access to the HPC resources of CALMIP under the allocation
2011-[P1115]. SOHO (MDI) is a mission of international cooperation between the European Space Agency
(ESA) and NASA. We are indebted to the Hinode team for the opportunity to use their data. Hinode
is a Japanese mission developed and launched by ISAS/JAXA, collaborating with NAOJ as a domestic partner,
NASA and STFC (UK) as inter-national partners.  Scientific operation of the Hinode mission is conducted by
the Hinode science team organized at ISAS/JAXA. This team mainly consists of scientists from institutes
in the partner countries.  Support for the post-launch operation is provided by JAXA and NAOJ (Japan),
STFC (UK), NASA, ESA, and NSC (Norway). The simulations were performed on the Pleiades supercomputer at
the Ames Research Center with resources provided by the High End Computing program through the NASA
Science Mission Directorate. We thank the anonymous referee for his/her careful reading of our manuscript and
his/her many insightful comments and suggestions.
\end{acknowledgements}

  \bibliographystyle{aa}    
\bibliography{biblio}

\end{document}